\documentstyle[11pt,aaspp4]{article}

\begin{document}
\title{DISCOVERY OF A BOXY PEANUT-SHAPED BULGE \\
IN THE NEAR INFRARED }

\author{
A. C. Quillen\altaffilmark{1}$^,$\altaffilmark{2}$^,$\altaffilmark{3},
L. E. Kuchinski$^{1,}$\altaffilmark{4},
J. A. Frogel$^{1,}$\altaffilmark{5}$^,$\altaffilmark{6}, \& D. L. DePoy$^{1,5}$
}

\altaffiltext{1}{Astronomy Department, Ohio State University, 174 W. 18th Ave., 
    Columbus, OH 43210}
\altaffiltext{2}{University of Arizona, Steward Observatory, Tucson, AZ 85721}
\altaffiltext{3}{E-mail: aquillen@as.arizona.edu}
\altaffiltext{4}{E-mail: lek@payne.mps.ohio-state.edu}
\altaffiltext{5}{Visiting Astronomer at Cerro Tololo Interamerican 
Observatories}
\altaffiltext{6}{also at Department of Physics, University of Durham, Durham, England}

\begin{abstract}
We report on the discovery of a boxy/peanut shaped bulge
in the highly inclined barred Seyfert 2 galaxy NGC~7582.  
The peanut shape is clearly evident in near infrared $JHK$
images but obscured by extinction from dust in visible $BVR$ images.
This suggests that near infrared imaging surveys will discover a larger
number of boxy/peanut morphologies than visible surveys, particularly
in galaxies with heavy extinction such as NGC~7582.
The bulge in NGC~7582 exhibits strong boxiness compared
to other boxy/peanut shaped bulges.
If the starburst was mediated by the bar, then it is likely
that the bar formed in less than a few bar rotation periods or
a few $\times 10^8$ years ago.
If the bar also caused the peanut, then
the peanut would have formed quickly; on a timescale of a few
bar rotation periods.

\end{abstract}


\section {INTRODUCTION}

	To date, the majority of surveys describing galaxy morphology 
have been carried out at visible wavelengths.  The advent of
large format infrared array detectors makes it possible 
to carry out such surveys in the near infrared and to search for 
stellar structures that at visible wavelengths would either be obscured by dust
or hidden by the presence of young stars.
Previous morphological studies in the near infrared have
discovered underlying stellar structures that are obscured by
dust.  For example, stellar bars have been discovered with near infrared 
images in NGC 1068 (\cite{sco88}) and M82 (\cite{tel91}) 
and in the nuclear regions of M100 (\cite{kna95}) and NGC 1097 (\cite{for92}).
These discoveries have been made typically in regions 
where high extinction 
and confusion from star formation makes it difficult to interpret the 
visible band images.  As a result near infrared imaging surveys find
a larger percentage of galaxies harbor bars 
than do visible imaging surveys (e.g. \cite{her96}).
Near-infrared images have also been used to probe the 
structure of highly inclined galaxies (e.g. \cite{kuc96})
where it is possible to study the vertical structure of the galaxy.
It is particularly difficult to identify underlying stellar structure in highly
inclined systems because 
even moderate amounts of dust can obscure structure in the plane
of the galaxy.

        The existence of spiral bulges with boxy or peanut-shaped morphology
has been recognized for many years (e. g. \cite{bur59}, \cite{dev74}).  
Identification of boxy or peanut-shaped bulge morphology has exclusively
been done with visible band images.  In galaxies with enhanced star
formation or large dust column depths boxy or peanut-shaped bulges
have been difficult to identify.
In this letter we report on the discovery in 
the near infrared wavelengths of a peanut-shaped
bulge in the Seyfert 2 galaxy NGC~7582.

Several mechanisms have been proposed to explain the
formation of boxy/peanut shaped
structures, including resonant heating by bars (\cite{com90}), 
bending instabilities
in the disk (\cite{rah91}; \cite{mer94}), and mergers or
accretion of small satellites (\cite{qui86}).
We note that these mechanisms involve kicking normal disk
population stars higher
above the plane of the galaxy, not forming new stars above the plane.
Recently, kinematic studies of spectra of a few edge-on 
peanut-shaped bulges have provided
strong evidence that these bulges are indeed barred 
(\cite{kui95}, \cite{bet94}).    In these few
galaxies it is likely that the bar caused the peanut.


\section{THE IMAGES OF NGC~7582}
NGC~7582 was observed in the
near infrared $J,H$ and $K$ bands and in the visible $B,V$ and $R$ bands
as part of a survey 
of $\sim 250$ galaxies  
currently being carried out at the Ohio State University.
The survey's goal is to produce a library of photometrically
calibrated images of late-type galaxies from $0.4$ to $2.2 \mu$m.
For notes on the observation and reduction techniques see
\cite{pog96}, or for individual examples \cite{qui94} and \cite{qui95}.
All the images were obtained 
at the Cerro Tololo Interamerican Observatories.
The galaxy was observed in $BVR$ 
at the 0.9m telescope on 1994 October 27 using the Tek\#2
$1024\times1024$ pixel CCD with a spatial scale of $0.40''$/pixel.
Total on source exposure times were 20, 15 and 10 minutes for $B,V$ and
$R$ respectively.  
The $JHK$ images were obtained at the 1.5m telescope 
on 1994 October 23 using a NICMOS 3 $256\times256$
pixel infrared array camera with a spatial scale of $1.16''$/pixel.
Total on source exposure times were 16.0, 15.0 and 29.2  minutes at $J,H$ 
and $K$ bands respectively.
All images were observed during clear but non-photometric conditions.
The visible images were calibrated with snapshots of the galaxy taken 
during photometric conditions 
on 1994 November 1 with the same camera and telescope.
The infrared images were calibrated on the CTIO/CIT system 
using aperture photometry of the galaxy from \cite{fro82}.

\section {RESULTS}

Figure 1 shows grayscale images of NGC~7582  overlayed with isophotes in
the $J$ and $V$ bands.
Figure 2 shows with $B-R$ and $V-H$ color maps.   
Extinction from dust is higher on the north-east side than on
the opposite side 
(see the $V-H$ color map).
The near infrared images of the bulge and disk are nearly symmetrical about
the galaxy's major axis.  They clearly show a boxy/peanut morphology for
the bulge of the galaxy
with boxy isophotes outside of peanut-shaped isophotes (see Figure 2).
The visible images, on the other hand, show no evidence for a
boxy or peanut
shape on the heavily obscured north-east side but peanut-shaped
isophotes can
be seen on the opposite side where there is less extinction.

NGC~7582 is a moderately inclined ($i \sim 65^\circ$) barred
galaxy with spiral arms visible outside the bar (see Figure 2a).
The galaxy is classified as SBSab in the {\it RC3} \cite{dev91}.
Most galaxies in which boxy/peanut-shaped bulges are seen have
such a high inclination that the presence of a bar is ambiguous.
This galaxy, like NGC~4442 \cite{bet94}, is remarkable in that
both the bar and boxy/peanut-shaped bulge can be observed
simultaneously.  
In moderately inclined galaxies 
where only one side of the galaxy is free of extinction 
such as NGC~7582, if peanut shaped isophotes are observed only
on one side, it is not possible to differentiate between
structure in the plane of the galaxy and a boxy/peanut shaped bulge.
The symmetry observed above and below the plane of the galaxy 
in the near infrared images makes it possible to clearly identify
the boxy/peanut in NGC~7582 as being part of the bulge and not structure
in the plane of the galaxy.

NGC~7582 is a Seyfert 2 galaxy (\cite{war80}) and a X-ray source
(\cite{ver81}).  It may  
have suffered an interaction with one of its two close
companions - NGC~7590 and NGC~7599 - which, together with NGC~7552
make up the Grus Galaxy Quartet.
The appearance of the bar suggests the presence of a considerable
amount of dust.
Its infrared flux, star formation rate and molecular gas mass
as traced in carbon monoxide are all higher than normal
(\cite{hec89}, \cite{cla92}).
The large gas mass ($M_{H_2} \sim 10^{10} M_\odot$; \cite{cla92}) is
associated with a large dust extinction, also 
inferred from the deep 10 micron
silicate absorption measured in front of its nucleus (\cite{fro82}).
The $J-K$ color increases by $\sim 0.1$ mag near the plane
of the galaxy compared to colors above and below this plane.
The peanut-shaped bulge was not previously recognized in visible images
because of this large extinction.

In the bulge (within $15''$ from the nucleus), 
colors more distant than $3''$ from the bar major axis
on the south-west side of the galaxy  (where there is little extinction)
are approximately constant in all
bands as a function of distance from this axis.
These colors are 
$J-K = 0.88$, $J-H = 0.62$, $B-V = 0.95$, $H-V = 2.75$, and $V-R = 0.56$
with errors of $\pm 0.08$ from photometric calibration.
These colors are similar to those of other Sb galaxies (\cite{fro88})
and arise from light from an older stellar population.
This would be consistent with the peanut formation theories mentioned
above where normal disk stars are kicked above the plane of the galaxy.
On the north-east side the colors are redder because of extinction
but regain the values of the south-west side for distances
greater than $20''$ from the bar major axis (outside the main
band of extinction). 
The near constant colors in the bulge itself are not inconsistent
with colors observed in other boxy/peanut shaped bulged galaxies 
(e.g.  \cite{sha93}).

We fit ellipses to the $K$ band isophotes using the ellipse routine
in the stsdas package of iraf which uses an 
iterative  method  described by \cite{jed87}.  The results
of this fit are shown in Figure 3.  The A4 or cos(4 $\theta$) amplitude
is divided by the semi-major axis and the local intensity gradient
and measures the isophote's deviations from perfect ellipticity.
A positive term represents diskiness
and a negative term boxiness.  Figure 3c shows that the bulge
is indeed quite boxy with deviations as large as $-5\%$.   The bulge
achieves peak boxiness at a semi-major axis of $\sim 22''$.
The boxy/peanut
galaxies studied in a similar fashion by  \cite{sha93} have a 
mean A4 of $-3.3\%$ with strongly boxed bulges at $\sim -4\%$.
The peak value we measure in NGC~7582 is perhaps slightly
stronger than the strongest boxy/peanuts seen \cite{sha93}'s sample
of galaxies.  

As mentioned in the introduction, many theoretical scenarios for
the formation of a boxy/peanut-shaped bulge require the presence
of a bar.  Bars can also cause starbursts by funneling gas into the center
of the galaxy.  
%
According to this scenario, the starburst begins either during
the gas inflow phase along
the bar as gas is concentrated into shocks aligned with the bar,
or in the nucleus (or nuclear ring) once the gas concentrated there 
reaches sufficiently high densities.  In either event
the starburst is expected to begin less than a couple of bar rotation 
periods after the bar has been formed (e.g. \cite{fri95}; \cite{hel94}).
The bar is about $140''$ long (or 14.7kpc using a  
Hubble constant of 75 km/s Mpc$^{-1}$ which gives a distance to NGC~7582 
of 21 Mpc).    For a circular velocity of 200 km/s (\cite{mor85})
the bar has a rotation period of $\sim 3 \times 10^8$ years.

It is likely that the 
the starburst in NGC~7582 was mediated by the bar.   
This implies that the bar formed recently.
Starbursts only last at most a few times $10^7$ years 
(e.g. Larson 1987, \cite{rie93}),  and bars only require
one or two bar rotation periods to concentrate gas, therefor
it is likely that the bar formed less than a few $\times 10^8$ years ago.

If the boxy peanut-shaped bulge (in addition to the starburst)
was caused by the bar 
then the boxy peanut also formed less than 
a few $\times 10^8$ years ago.   
However, predicted timescales for creating a boxy bulge range
from the order of 10 bar rotation periods 
for resonant heating 
(\cite{com90}; \cite{pfe90})  to a few bar
rotation periods for the bending or firehose instability 
(\cite{rah91} ; \cite{mer94}).
Because we suspect that the peanut shape has formed recently
in NGC~7582, it is reasonable to suggest that it formed by one of
the faster proposed mechanisms.

%

We expect that near infrared imaging surveys will discover more
boxy/peanut-shaped
bulges, analogous to the larger percentage of bars identified in near infrared
surveys than visible ones (e.g. \cite{her96}).
All existing surveys designed to detect boxy or peanut-shaped
bulges have been carried out at optical wavelengths.
These surveys find that a substantial fraction of galaxies may
have boxy/peanut-shaped bulge morphologies
(\cite{sha87}, \cite{det89}, \cite{jar86}, and \cite{dd87}); for example,
\cite{sha87} finds that $20\%$ of highly inclined galaxies
in the RC2 exhibit boxy or peanut-shaped bulges on the
POSS B and R or the ESO/SERC J survey plates.
Surveys that include near infrared as well as visible data will provide better
statistics on the frequencies of both bars and boxy/peanut-shaped
bulges.   In moderately inclined systems such as NGC~7582 more systems
with both bars and peanuts will be identified.
These statistics should determine if the two
features are commonly related. 

As the mechanisms for boxy/peanut bulge formation show, bars can 
provide a major source of stellar heating (as in random motions of stars).
Studies which determine 
the fraction of galaxies with bars and boxy/peanut shaped bulges and the
connection between bars and boxy/peanut shaped bulges should
shed light on the role of bars on secular evolution in disk galaxies.  
As in the case of NGC~7582, boxy/peanuts bulges are more 
likely to be discovered in infrared surveys in galaxies 
with large dust extinction and active star formation.
By investigating systems that feature both boxy/peanut bulges and
short-lived phenomena such as starbursts, it may be possible to place 
limits on the timescale for vertical heating in barred galaxies.



\acknowledgments

We thank the referee for constructive criticism which resulted in
a much clearer paper.
We are grateful to Roberto Aviles for obtaining the $B,V$ and $R$
images for us.
We acknowledge helpful discussions and correspondence with G. Rieke.
The OSU galaxy survey is being supported in part by NSF grant AST 92-17716.
A.C.Q. acknowledges the support of a Columbus fellowship.
J.A.F. thanks
Roger Davies for his hospitality at Durham University and PPARC for
partial support via a Visiting Senior Research Fellowship.

\clearpage

\begin{figure*}
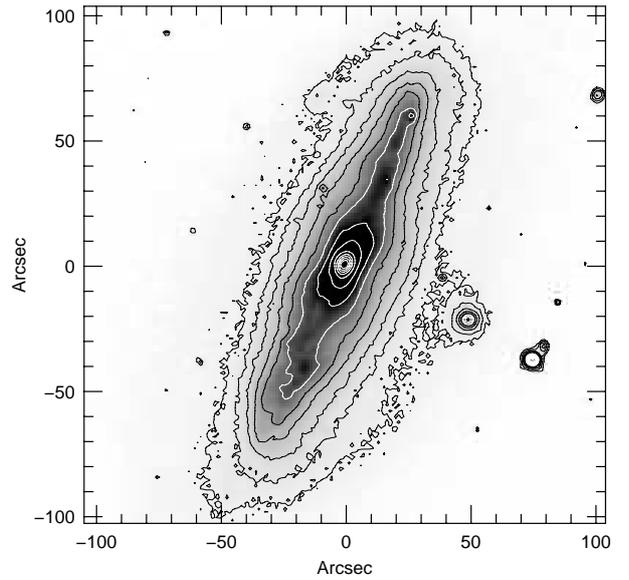

\caption[junk]{
a) $V$ band greyscale image of NGC~7582.
North is up and east is to the left.
b) $J$ band greyscale image.
c) $V$ band greyscale image of NGC~7582 overlayed with isophotes.
Contours are displayed from 17.0 to  23.0 mag/arcsec$^2$ and are 0.5 mag apart.
The peanut shaped bulge is completely obscured on the north-east
side of the galaxy, though peanut shaped isophotes can be
seen on the opposite side.
d) $J$ band greyscale image overlayed with isophotes.
Contours are displayed from 14.5 to  20.5 mag/arcsec$^2$ and are 0.5 mag apart.
Boxy-shaped isophotes outside peanut-shaped isophotes are clearly visible
in this image.
\label{fig:fig1} }
\end{figure*}

\begin{figure*}
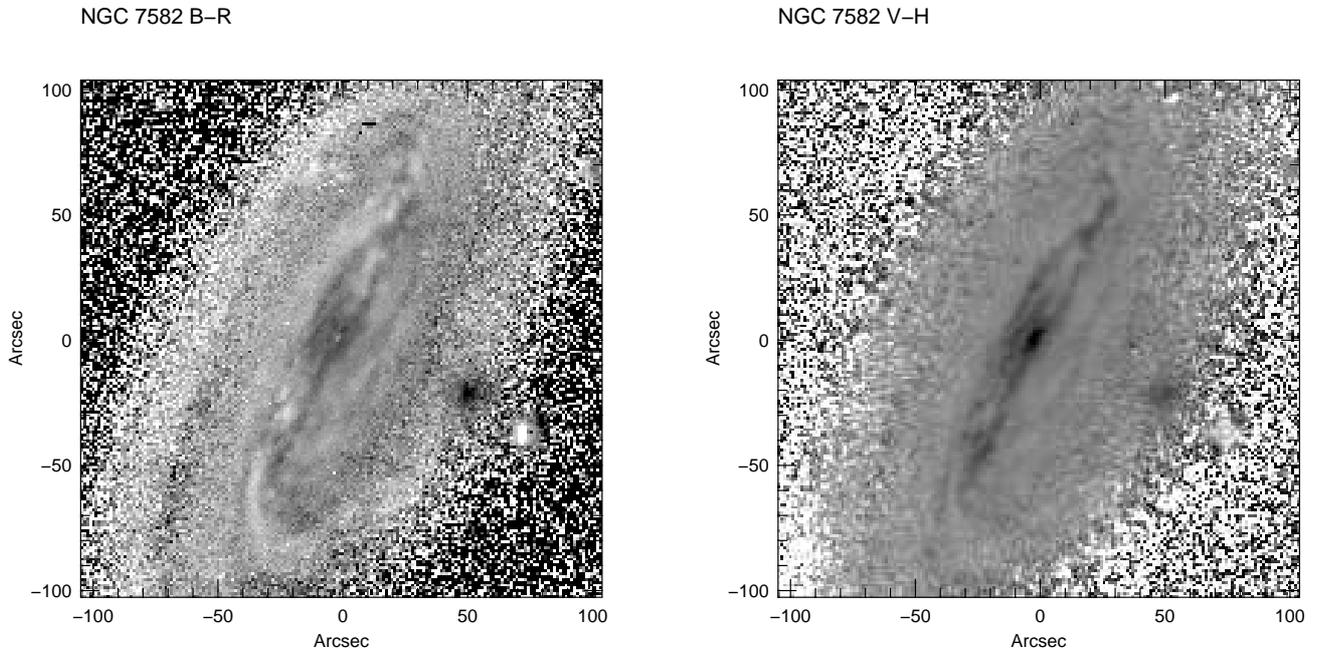

\caption[junk]{
a) $B-R$ color map.  Darker areas are redder than lighter areas.
b) $V-H$ color map.  
Extinction is higher on the north-east side of the galaxy
than on the opposite side.
\label{fig:fig2} }
\end{figure*}

\begin{figure*}
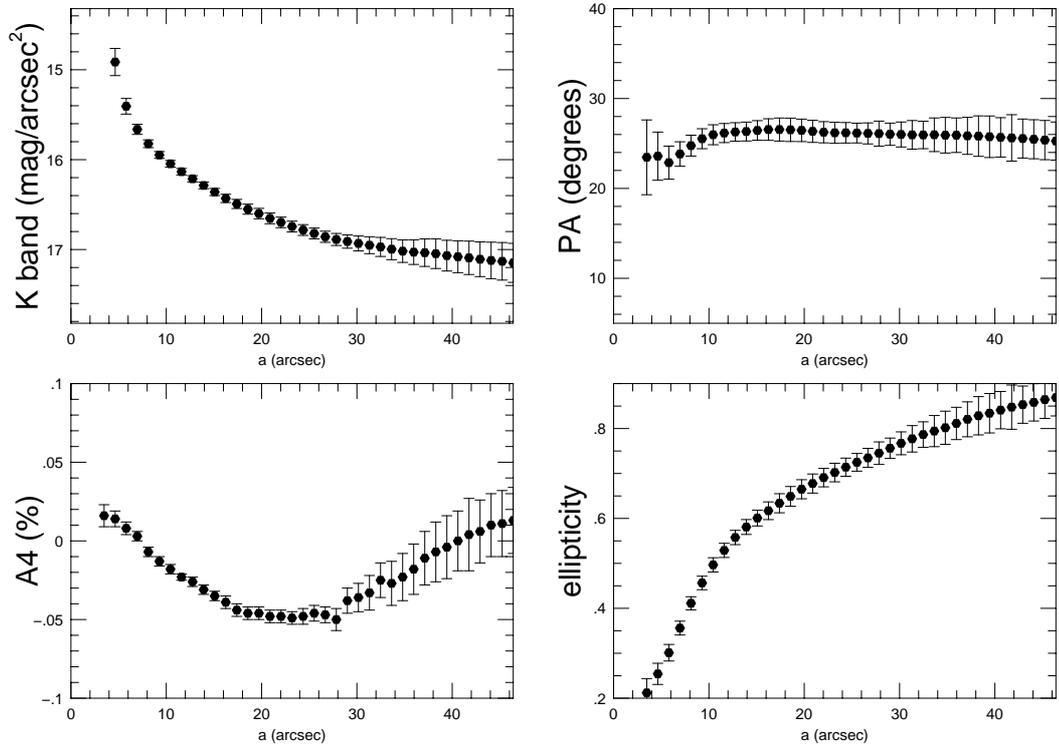

\caption[junk]{
Results of ellipse fitting to the K band isophotes of NGC~7582.
The horizontal axis in all plots gives the semi-major axis in arcsecs.
a) Surface brightness at K band.
b) Position angle of ellipses.
c) Cos(4 $\theta$) component.
A negative value represents boxiness whereas a positive value
represents diskiness.
b) Ellipticity.
\label{fig:fig3} }
\end{figure*}

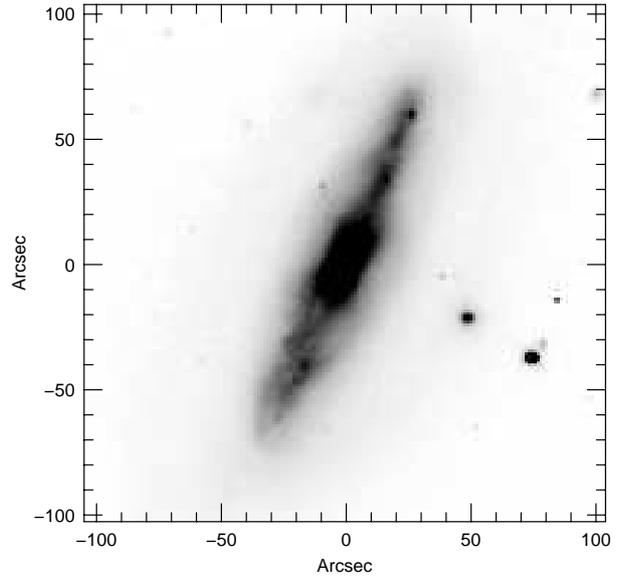
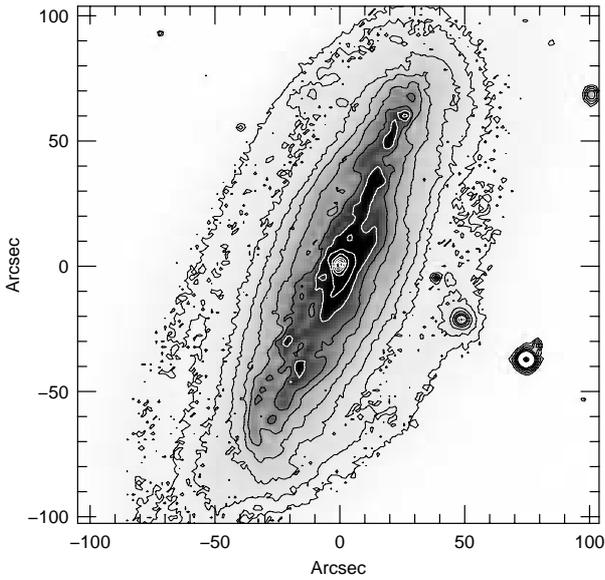
\end{document}